\documentclass[final,5p,twocolumn,times]{elsarticle}

\usepackage{longtable}
\usepackage{amssymb}
\usepackage{amsthm}
\usepackage{dcolumn}
\usepackage{graphicx}
\biboptions{sort&compress}

\journal{Journal of Magnetism and Magnetic Materials}

\newcommand{\dto}{$\rm Dy_2Ti_2O_7$}
\newcommand{\yto}{$\rm Y_2Ti_2O_7$}
\newcommand{\hto}{$\rm Ho_2Ti_2O_7$}
\newcommand{\dyto}{$\rm DyYTi_2O_7$}
\newcommand{\hyto}{$\rm HoYTi_2O_7$}
\newcommand{\rto}{$R_2$Ti$_2$O$_7$}
\newcommand{\ryto}{$R$YTi$_2$O$_7$}

\newcommand{\dy}{$\rm Dy^{3+}$}
\newcommand{\ho}{$\rm Ho^{3+}$}
\newcommand{\y}{$\rm Y^{3+}$}

\begin{document}

\begin{frontmatter}

\title{Heat transport of the spin-ice materials \hto\ and \dto}

\author{S. Scharffe}
\author{G. Kolland}
\author{M. Valldor\fnref{fn1}}
\author{V. Cho}
\author{J.F. Welter}
\author{T. Lorenz \corref{cor1}}

\ead{tl@ph2.uni-koeln.de}

\cortext[cor1]{Corresponding author}

\address{Physikalisches Institut, Universit\"{a}t zu K\"{o}ln, Z\"{u}lpicher Str. 77, 50937 K\"{o}ln, Germany}

\fntext[fn1]{Present address: Max-Planck-Institut f\"{u}r Chemische Physik fester Stoffe, Noethnitzer Strasse 40, 01187 Dresden, Germany}

\begin{abstract}

The elementary excitations of the spin-ice materials \hto\ and \dto\ in zero field can be described as independent magnetic monopoles. We investigate the influence of these exotic excitations on the heat transport by measuring the magnetic-field dependent thermal conductivity $\kappa$. Additional measurements on the highly dilute reference compounds \hyto\ and \dyto\ enable us to separate $\kappa$ into a sum of phononic ($\kappa_{\rm{ph}}$) and magnetic ($\kappa_{\rm{mag}}$) contributions. For both spin-ice materials, we derive significant zero-field contributions $\kappa_{\rm{mag}}$, which are rapidly suppressed in finite magnetic fields. Moreover, $\kappa_{\rm{mag}}$ sensitively depends on the scattering of phonons by magnetic excitations, which is rather different for the Ho- and the Dy-based materials and, as a further consequence, the respective magnetic-field dependent changes $\kappa_{\rm{ph}}(B)$ are even of opposite signs. 
\end{abstract}

\end{frontmatter}

\section{Introduction}
\label{sec:Intro}
The spin-ice materials \hto\ and \dto\ are continuously attracting lots of attention due to their residual ground-state entropy and anomalous low-energy excitations, which can be described as magnetic monopoles\cite{Castelnovo2008}. Both materials crystallize in the cubic pyrochlore structure, where the magnetic \ho\ or \dy\ ions form a network of corner-sharing tetrahedra. The (2J+1)-fold degeneracy of the single-ion's Hund's rule ground state with total momentum $J=L+S$ is lifted by the crystal electric field (CEF) and the lowest-lying sublevel is a doublet, which almost completely consists of the  $\left| \pm J_z^{\rm{max}} \right\rangle$ state with $J_z^{\rm{max}}=8 \: (15/2)$ for the Ho(Dy)-based material~\cite{Malkin2010,Jana2002,Kitagawa2008}. For each ion, the local quantization axis points from the corner to the center of the tetrahedron, {\it i.e.} along one of the $\left\{ 111 \right\}$ directions of the cubic structure. Because the energy difference to the first excited sublevels is of the order of 200--300~K, the low-temperature magnetism of both materials can be well described by non-collinear $S=1/2$ Ising spins with large magnetic moments of about 10~$\mu_{\rm B}$. Antiferromagnetic exchange interactions are so weak that the dipolar interactions dominate, which favor a six-fold degenerate groundstate with two spins pointing into and two out of each tetrahedron. This "2in/2out" arrangement is equivalent to Pauling's ice rule describing the hydrogen displacement in water ice and results in a residual zero-temperature entropy\cite{Nagle1966,Ramirez1999,Bramwell2001,Hiroi2003,Sakakibara2003}. Flipping a single spin creates a pair of "3in/1out" and "1in/3out" excitations on neighboring tetrahedra and due to the ground-state degeneracy such a pair fractionalizes into two individual excitations that can freely propagate through the crystal and can be described as independent magnetic (anti-)monopoles. The dynamics of these anomalous excitations is subject of intense research~\cite{Ryzhkin2005,Morris2009,Giblin2011,Castelnovo2011,PhysRevLett.105.267205,Kadowaki2009,Jaubert2011,Bramwell2009,Blundell2012,Matsuhira2011,Yaraskavitch2012,Grams2013}. 

In the present work, we discuss the influence of such magnetic monopole excitations on the heat transport of spin-ice materials. As \hto\ and \dto\ are good insulators, the heat transport is dominated by phonons and the magnetic excitations may influence the total heat transport in two ways. The magnetic excitations might add an additional contribution to the heat transport or they scatter with phonons and therefore suppress the phonon heat transport. In general, both effects are present and as an approximate Ansatz the superposition $\kappa \simeq \kappa_{\rm{ph}}+\kappa_{\rm{mag}}$ can be used, where both individual contributions $\kappa_{\rm{ph}}$ and $\kappa_{\rm{mag}}$ are reduced compared to their hypothetical bare values by phonon-magnon scattering. Concerning the above-described monopole excitations in spin ice, we are not aware about any prediction of the expected magnitude of $\kappa_{\rm{mag}}$. On the one hand, their typical energy scale is low, while, on the other hand, their mobility in zero magnetic field is high. Moreover, these excitations are not described by a standard quasi-particle dispersion, which might be the most important issue of their dynamics. In this respect, some similarities between the monopole/antimonopole excitations and the two-spinon continua of one-dimensional quantum spin chains can be expected. During the last years, intense studies of the (magnetic) heat transport of low-dimensional quantum spin systems have been performed, but still many aspects are not yet understood~\cite{Hess2007,Sologubenko2007}.

For low-dimensional spin systems, the expected anisotropy of $\kappa_{\rm{mag}}$ is typically used to separate it from $\kappa_{\rm{ph}}$. This strategy is not possible in the three-dimensional spin-ice materials, but weak magnetic fields in the range of less than 1~T are sufficient to lift the ground-state degeneracy and thus to prevent the monopole/antimonopole deconfinement and somewhat larger fields even cause a full saturation of  the magnetization, meaning that any types of magnetic excitations can be strongly suppressed due to the large Zeeman splitting. Another strategy to separate $\kappa_{\rm{ph}}$ and $\kappa_{\rm{mag}}$ is to study (non-)magnetic reference compounds of the same structure. For the spin-ice materials this can be achieved by the substitution series $\rm (Dy_{1-x}Y_x)_2Ti_2O_7$ and $\rm (Ho_{1-x}Y_x)_2Ti_2O_7$ because of the very similar ionic radii of \dy, \ho, and the non-magnetic \y.   

The magnetic-field dependent heat transport of \dto\ has been studied by different groups and a significant decrease of $\kappa(B)$ in the low-temperature range is observed~\cite{Klemke2011,Sun2013,Kolland2012,Kolland2013,Scharffe2013}. Refs.~\cite{Klemke2011,Sun2013} assume a purely phononic heat transport in zero field and a field-induced suppression of $\kappa_{\rm{ph}}$ by some field-dependent scattering mechanisms which is not further specified. This differs from our interpretation~\cite{Kolland2012,Kolland2013,Scharffe2013}, which is based on the comparative study of $\rm (Dy_{1-x}Y_x)_2Ti_2O_7$ for $x=0$, 0.5, and 1. Our data reveal that, in the field range above about 1.5~T, a very similar Dy-related, field-induced suppression of $\kappa_{\rm{ph}}(B)$ is present for both, the spin ice \dto\ and  the highly dilute \dyto, which does not show spin-ice behavior. In \dto, however, we observe an additional low-field dependence of $\kappa(B)$, whose anisotropic field dependence and hysteresis behavior clearly correlates with the spin-ice physics. This evidences a sizeable $\kappa_{\rm{mag}}$ in zero field, which is successively suppressed by the application of a magnetic field due to the field-induced suppression of the monopole mobility. Experimental evidence for a zero-field monopole contribution to the heat transport has been also proposed from an analysis of  $\kappa(B,T)$ of \hto \cite{Toews2013}. However, the magnitude of $\kappa_{\rm{mag}}$ estimated for \hto\ is more than an order of magnitude smaller than  $\kappa_{\rm{mag}}$ of \dto. Moreover, above about 0.6~K the overall field dependence $\kappa(B)$ of \hto\ is of the opposite sign than that of \dto. These strong differences motivated us to perform a more detailed comparative study of the field-dependent heat transport in \hto, \dto\ and the corresponding reference materials \hyto\ and \dyto.

\section{Experimental}
\label{sec:exp}

Single crystals of ($R_{1-x}$Y$_x$)$_2$Ti$_2$O$_7$ with $R={\rm Dy}$, Ho were grown by the floating-zone technique in a mirror furnace. The measurements of the thermal conductivity $\kappa(B)$, the magnetization $M(B)$ and the magnetostriction $\Delta L(B)/L_0$ were performed on oriented crystals of approximate dimensions $3 \times 1 \times 1 \, {\rm{mm}}^3$. Details of the sample preparation and the measurement techniques are given in Refs.~\cite{Kolland2012,Kolland2013,Scharffe2013}. Demagnetization effects were taken into account and the internal magnetic field was calculated for all measurements. Here, we mainly restrict to longitudinal configurations, that is, we measured the length changes $L\|B$ and, in most cases, applied the heat current $j$ along the longest sample dimension in order to minimize demagnetization effects. The only exceptions are the $\kappa(B\| [100])$ measurements of both Dy-based materials from Ref.~\cite{Kolland2013}, which were measured with $j\| [011]$ and $B\perp j$. These different configurations can influence the absolute values of $\kappa$, but we checked on $\rm (Dy_{1-x}Y_x)_2Ti_2O_7$ (with a different $x$)  that such differences are irrelevant for the following discussion.

\section{Results and Discussion}
\label{sec:Res_Dis}

\begin{figure}[t]
    \includegraphics[width=\linewidth]{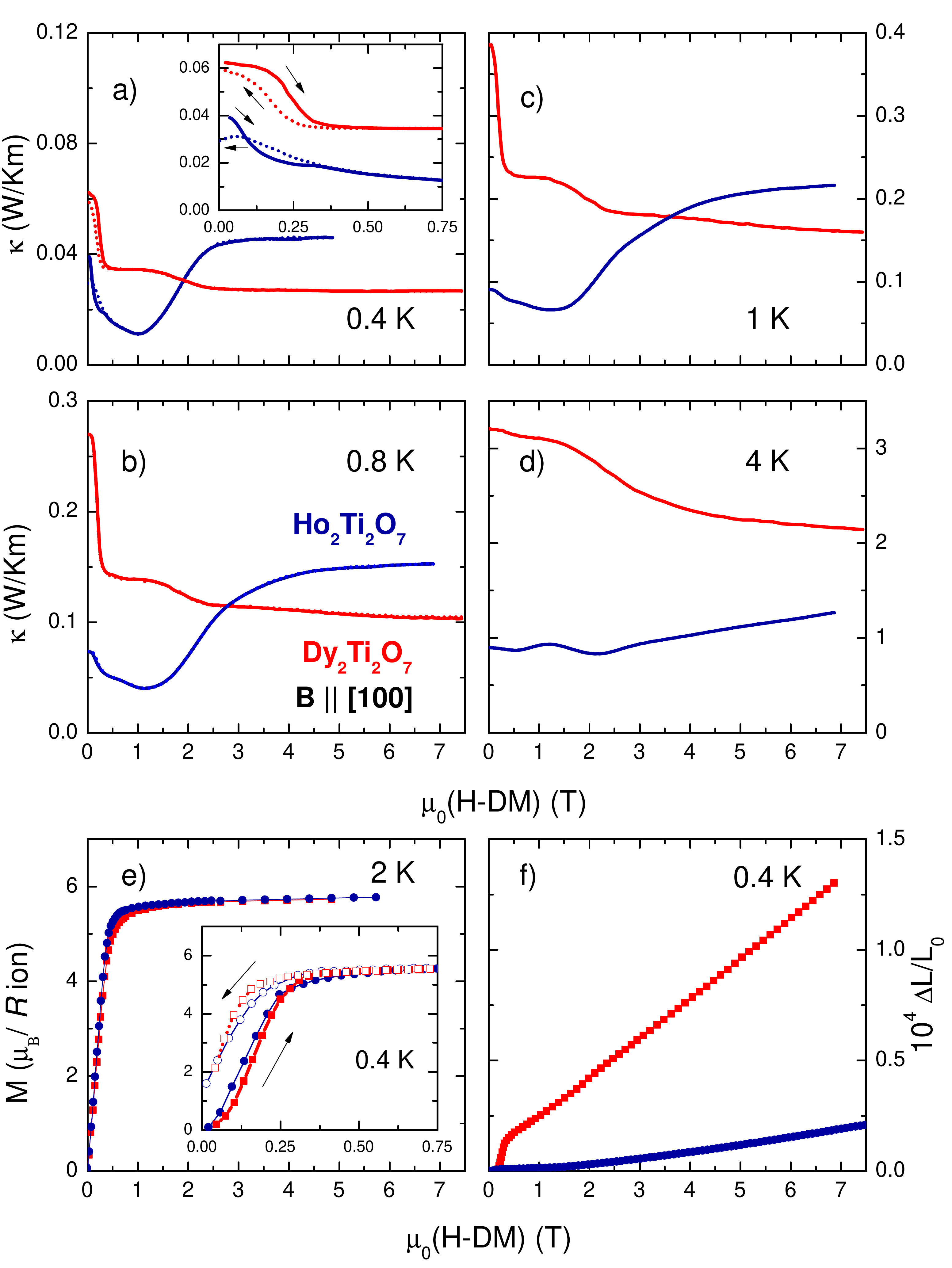}
    \caption{(Color online) (a--d): Thermal conductivity $\kappa (B)$ of \dto\ (red) and \hto\ (blue) as a function of the magnetic field $B\|[100]$ for different temperatures. The inset of panel (a) magnifies the low-field range. Panels (e) and (f) show corresponding measurements of the magnetization $M(B)$ with $R={\rm Ho}$ or Dy, and the magnetostriction $\Delta L(B)/L_0$. All the data were measured with increasing and decreasing magnetic field, but hysteresis effects only occur below about 0.6~K, as can be examplarily seen in the insets. \label{fig:1}}
  \end{figure}

\begin{figure}[t]
    \includegraphics[width=\linewidth]{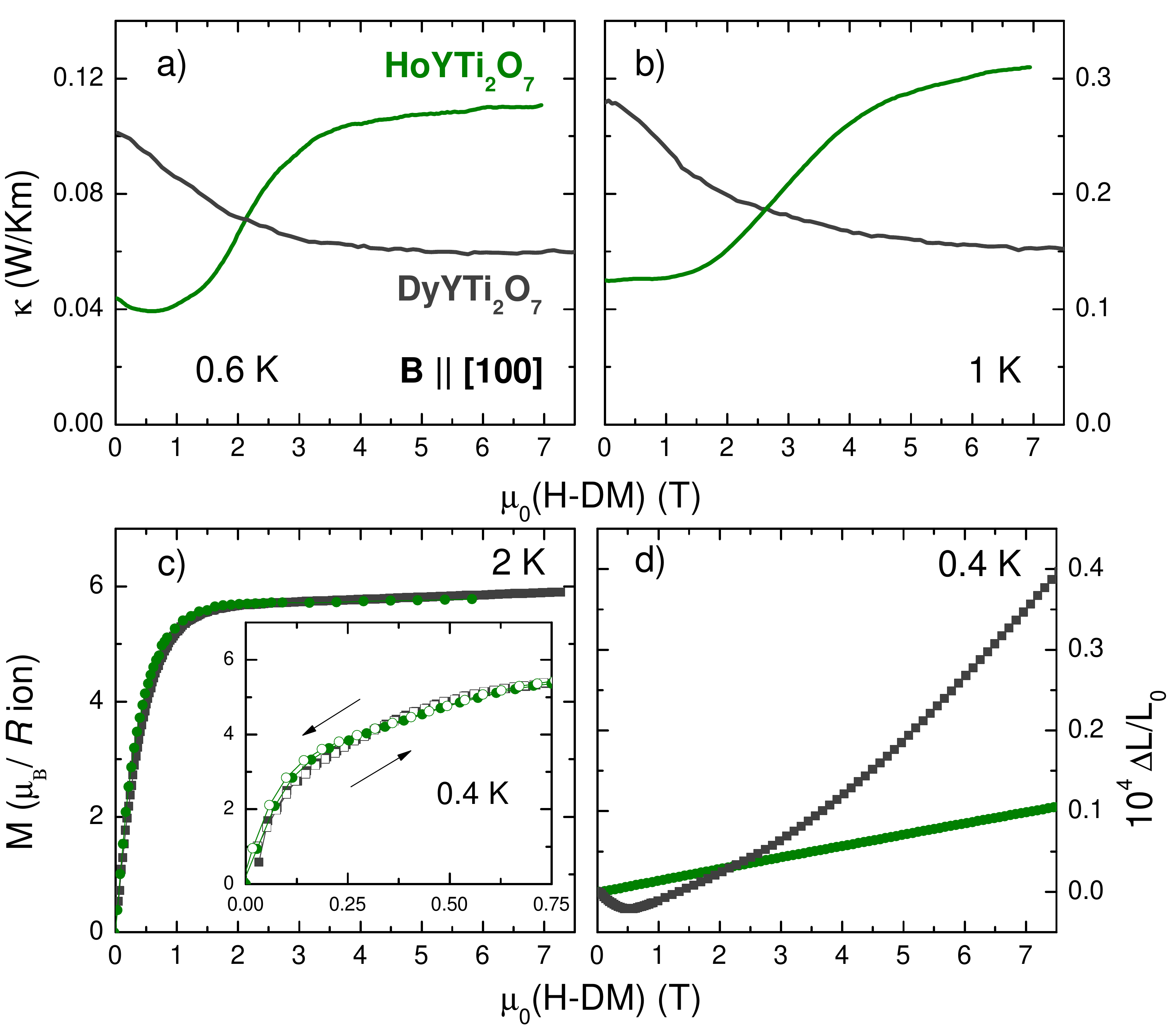}
    \caption{(Color online) Thermal conductivity (a,b) along the $[100]$ direction, magnetization (c), and magnetostriction (d) of \ryto\ with $R={\rm Dy}$ (dark grey) and $R={\rm Ho}$  (green) as a function of the magnetic field $B\|[100]$ for different temperatures. All the data were measured with increasing and decreasing field, but even at 0.4~K no sizeable hysteresis is present (see inset). \label{fig:2}}
  \end{figure}

\begin{figure}[t]
    \includegraphics[width=\linewidth]{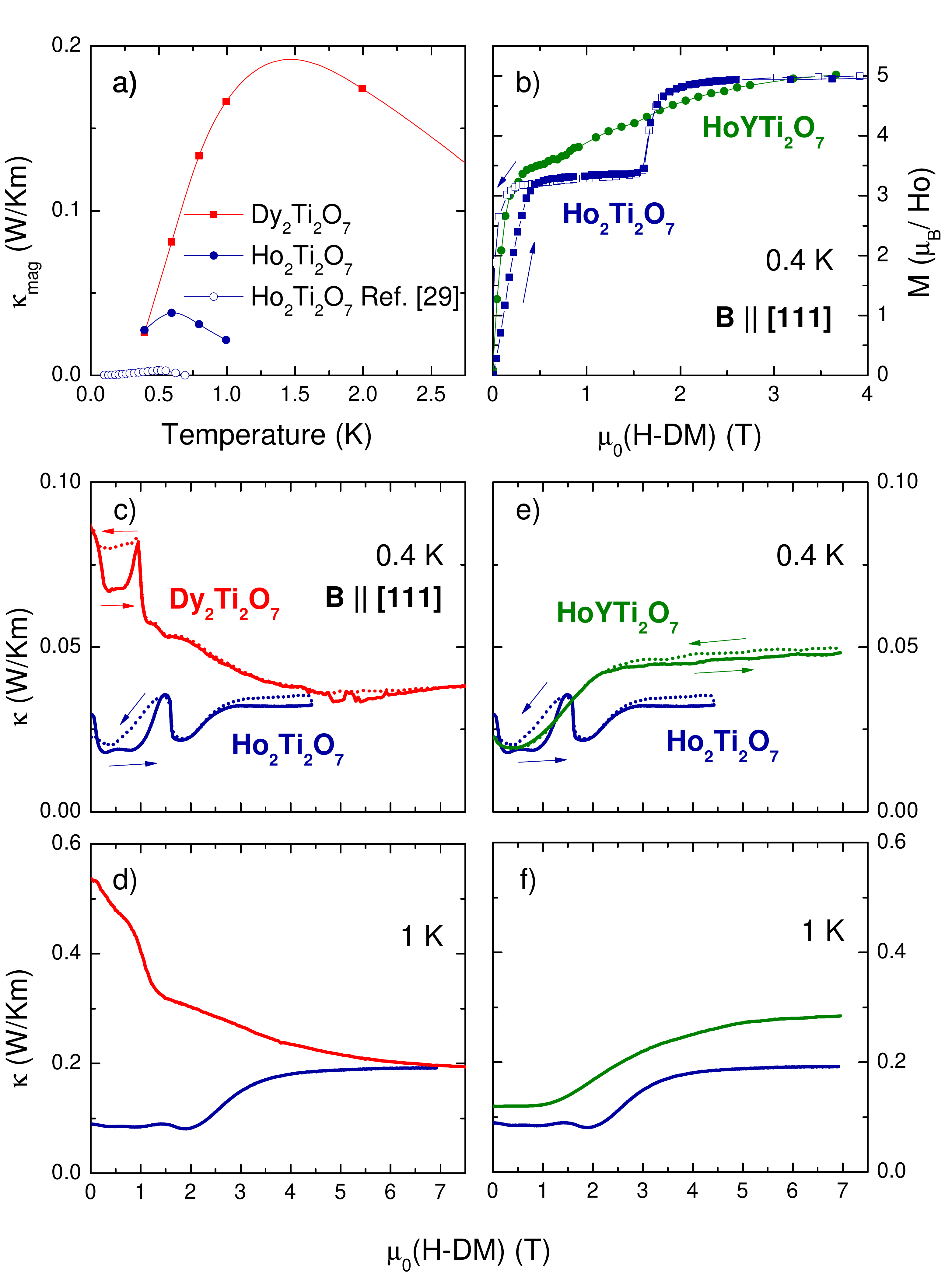}
    \caption{(Color online) (a): Magnetic contribution $\kappa_{\rm{mag}}$ of the zero-field heat transport of \hto\ and \dto. (b): Magnetization of \hto\ and \hyto\ for $B\| [111]$. (c--f): Comparison of the thermal conductivity $\kappa(B)$ of the spin-ice materials \hto , \dto\ and  the reference material \hyto.  \label{fig:4}}
  \end{figure}

Fig.~\ref{fig:1} compares representative measurements of $\kappa(B)$ for both spin-ice materials in the temperature range from 0.4 to 4~K, which clearly reveal that the overall field dependences of $\kappa(B)$ in the higher field range are very different. \dto\ shows a continuous decrease of $\kappa(B)$ above about 1.5~T, whereas $\kappa(B)$ of \hto\ increases with magnetic field. As is seen  in Fig.~\ref{fig:1}(e), the magnetization of both materials is essentially saturated above about 1~T in this low-temperature range. The opposite field dependences of $\kappa(B)$ are not  related to the spin-ice physics, which mainly takes place below 1~T. In the field range below about 0.5~T, $\kappa(B)$ of both materials shows a rapid drop, which can be attributed to a field-induced suppression of $\kappa_{\rm{mag}}(B)$. In particular towards higher temperature, this effect is significantly less pronounced in \hto\ than it is in \dto. On this qualitative level, the data of Fig.~\ref{fig:1} seem to confirm the previous result of Ref.~\cite{Toews2013}, but the latter was measured for different directions of the heat current and the applied magnetic field, namely $j\|B\| [111]$. In order to get a more quantitative estimate of $\kappa_{\rm{mag}}(B)$, the field dependence of the underlying phononic contribution $\kappa_{\rm{ph}}(B)$ in the low-field range is required. Therefore, we also studied $\kappa(B)$ of the highly dilute materials \dyto\ and \hyto. As half of the magnetic ions are replaced by non-magnetic \y , the spin-ice physics is expected to be essentially suppressed. This is confirmed by the magnetization data for $B\| [100]$ that do not show any hysteresis even at $T=0.4$~K, in contrast to the corresponding $M(B)$ curves of the spin-ice materials; compare the insets of Fig.~\ref{fig:1}(e) and Fig.~\ref{fig:2}(c). Moreover, the characteristic kagom\'{e}-ice plateau in the  $M(B)$ curves for $B\| [111]$ has vanished in the dilute material; see Fig.~\ref{fig:4}(b).    

In Fig.~\ref{fig:2}, characteristic $\kappa(B)$ measurements of \dyto\ and \hyto\ are compared. As the spin-ice physics in these materials is almost completely suppressed, these data yield clear evidence that the opposite field dependences of $\kappa(B)$ of the Dy- and the Ho-based materials arise from different field dependences of the phononic background. This raises the question why $\kappa_{\rm{ph}}(B)$ decreases with increasing field in \dyto\ and increases in \hyto. As described above, localized magnetic moments may serve as scattering centers for the phonons and because spin flips are suppressed in large magnetic fields, this mechanism can explain an increase of $\kappa_{\rm{ph}}(B)$ as it is observed in \hyto, but it cannot explain the decreasing $\kappa_{\rm{ph}}(B)$ of \dyto. As we have already discussed in Ref.~\cite{Kolland2013}, the decrease of $\kappa_{\rm{ph}}(B)$ is probably related to magnetic-field induced lattice distortions, which arise from the fact that the local quantization axes of the magnetic ions at the different corners of each tetrahedron are not collinear. Consequently, for any direction of the external magnetic field at least 3/4 of the magnetic ions feel a symmetry-breaking transverse field component, which mixes the higher-lying levels into the groundstate doublet, which in zero field almost completely consists of the  $\pm J_z^{\rm{max}}$ state~\cite{Malkin2010,Jana2002,Kitagawa2008}. This effect causes a van Vleck susceptibility, which is seen as a finite positive slope of the high-field magnetization data that varies between about 0.3 to 1\%/T, depending on the sample and the  field direction~\cite{Kolland2013,Prabhakaran2011,Fukazawa2002,Flood1974}. A further consequence is a pronounced anisotropic magnetostriction, {\it i.e.} field-induced length changes $\Delta L_i(B)$. In \dto, we found a significant elongation of $\Delta L_i\| B$ and weak contractions of $\Delta L_i\perp B$~\cite{Kolland2013}. Within a simplified classical picture, such field-induced lattice distortions may be visualized as resulting from the finite torques $\vec{\mu} \times \vec{B}$, which tend to align the non-collinear localized magnetic moments towards the field direction. With respect to the phononic heat transport, $\kappa_{\rm{ph}}(B)$ may decrease with increasing field due to the reduced lattice symmetry. In addition, the spin-flip rate may also increase due to the stronger mixing of the $\pm J_z^{\rm{max}}$ states with other $J_z$ levels, but this effect should vanish towards larger fields when spin flips are suppressed by the enhanced Zeeman splitting. Summarizing the discussion so far, there are different mechanisms which may either increase or decrease $\kappa_{\rm{ph}}(B)$ and it is difficult to predict which of them dominates. Experimentally, we find that the Dy- and the Ho-based materials are very different in this respect and this difference is not restricted to $\kappa_{\rm{ph}}(B)$. As is shown in Figs.~\ref{fig:1}(f) and~\ref{fig:2}(d), the magnetostriction for both  Dy-based materials is about 4 times larger than that of the corresponding Ho-based ones. This suggests that the stronger magnetostriction in the Dy-based materials seems to 
make the distortion-induced decrease of $\kappa_{\rm{ph}}(B)$ the dominant process, whereas in the Ho-based materials the decreasing phonon scattering by spin flips is dominant.

\begin{figure}[t]
    \includegraphics[width=\linewidth]{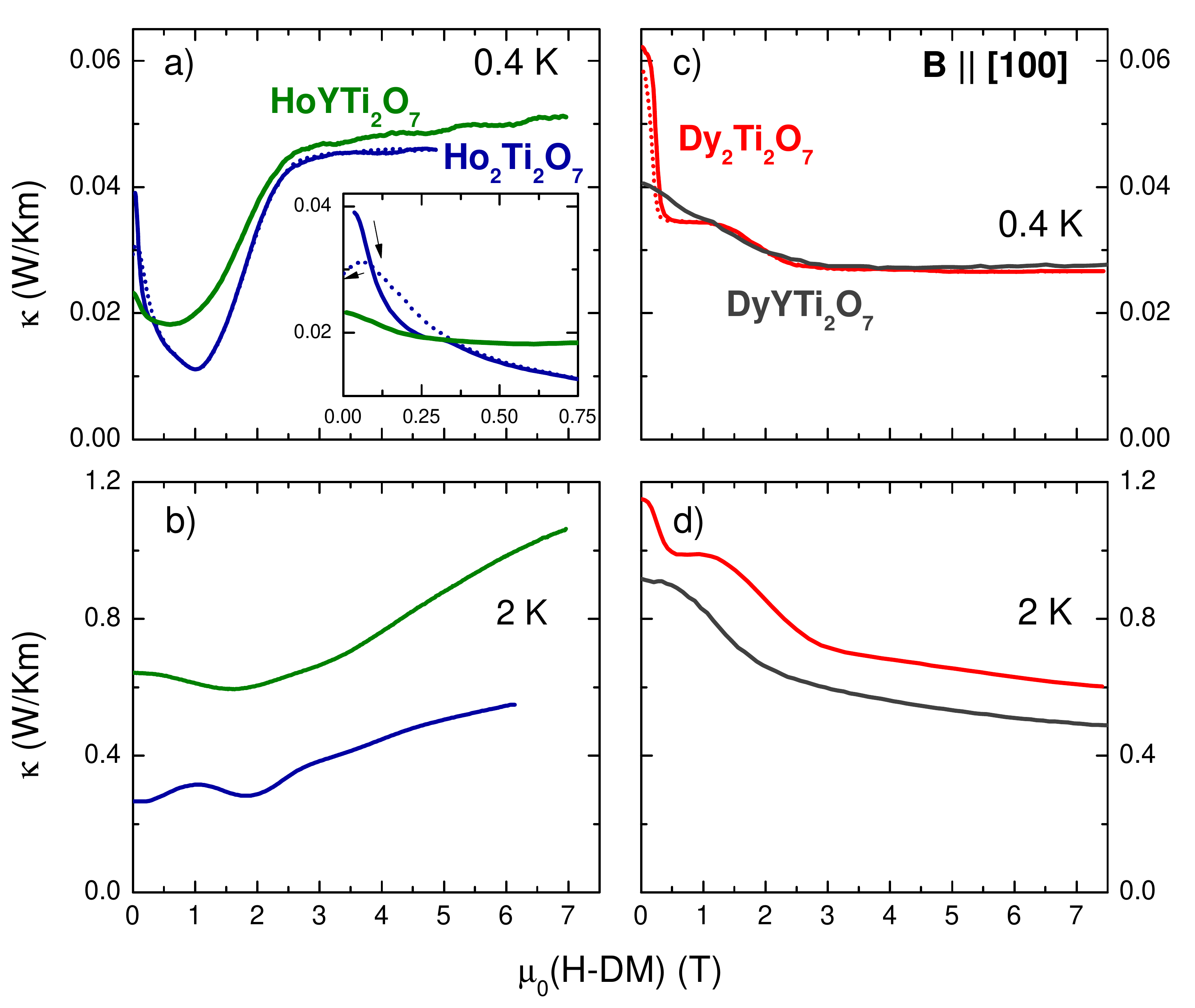}
    \caption{(Color online) Comparison of the thermal conductivity $\kappa(B)$ of the Ho-based (left) and the Dy-based (right) spin-ice materials \rto\ with the corresponding $\kappa(B)$ of the respective non-spin-ice reference materials \ryto. The inset of panel (a) shows an expanded view of the low-field range.\label{fig:3}}
  \end{figure}

Because the electronic configurations of \dy\ and \ho\ just differ by one electron ($4f^9$ {\it vs.} $4f^{10}$) in the inner $4f$ shell, it may appear surprising that the magnetostriction $\Delta L(B)$ and the magnetic-field dependent phonon heat transport $\kappa_{\rm{ph}}(B)$ of the Dy- and the Ho-based materials are so different. However, both quantities depend on various material parameters and, in  particular, the differences in the crystal-field level schemes of both ions may become important~\cite{Benito2004,Malkin2010,Jana2002,Kitagawa2008}, {\it e.g.} the fact that there is a Kramers protection of the zero-field doublet states of \dy\ but not for those of \ho. With respect to the question of a possible heat transport via magnetic monopoles, the different $\kappa_{\rm{ph}}(B)$ in the higher-field range are of minor importance because the spin-ice behavior is restricted to low fields. Therefore, the main question in this context is, whether it is possible to obtain a reliable estimate of the phononic background. 

In Fig.~\ref{fig:3} we directly compare representative $\kappa(B)$ measurements of the spin-ice materials \hto\ and \dto\ with $\kappa(B)$ of \hyto\ and \dyto.  Unfortunately, it is not possible to simply consider the difference between the $\kappa(B,T)$ curves of the pure and the respective reference material. The main reason is that the absolute values of $\kappa$ of different samples differ, which is partly due to experimental errors as, {\it e.g.}, the exact determination of the sample's geometry. This uncertainty should not exceed 20\% and could be treated by a temperature- and field-independent scaling factor. More important for a transport property is, however, its dependence on defect and impurity scattering. Because $\kappa$ usually increases with increasing sample quality, one may expect somewhat lower values of $\kappa$ for the dilute reference compound than for the pure spin-ice. This is more or less fulfilled for the Dy-based materials, but not for the Ho-based ones. If, however, the above-described spin-flip scattering is a dominant scattering mechanism for $\kappa_{\rm{ph}}$ in a certain temperature and field range, its decrease due to the lower Ho content may overcompensate an increasing Ho/Y-disorder scattering in the dilute material. In addition, $\kappa_{\rm{ph}}$ can be reduced by scattering via crystal-field excitations of the partially filled $4f$ shells of Ho and Dy, but due to the rather large energy splitting this effect should become relevant towards higher temperatures and, indeed, $\kappa(T)$ of \yto\ significantly exceeds that of \dto\ in the range of about 2 to 100~K~\cite{Kolland2012}. Due to all these reasons and the possibility that even the half-doped materials may still show some remnant spin-ice behavior we have to conclude that an unambiguous quantitative determination of the phonon background $\kappa_{\rm{ph}}(B)$ is not possible. As already discussed in Ref.~\cite{Kolland2013}, however, it appears reasonable to assume an essentially field-independent $\kappa_{\rm{ph}}^{B\rightarrow 0}$ in the low-field range.  Because $\kappa(B\|[100])$ of \dto\ shows a step-like decrease to a pronounced plateau around 1~T, which  anticorrelates with the rapid saturation of the magnetization for this field direction, we estimated $\kappa_{\rm{ph}}^{B\rightarrow 0}$ by these plateau values and derived the magnetic heat transport  for different field directions via $\kappa_{\rm{mag}}(B)\simeq \kappa(B)- \kappa_{\rm{ph}}^{B\rightarrow 0}$. From this analysis, which has to be restricted to the low-field range below about 1~T, we derived in Ref.~\cite{Kolland2013} an anisotropic $\kappa_{\rm{mag}}(B)$ whose magnitude reflects the different degeneracies $D$ of the various magnetic-field induced spin-ice groundstates. The maximum $\kappa_{\rm{mag}}$ is present in the zero-field state with $D=6$ and it is completely suppressed in the fully polarized states with $D=0$ for $B^{\|100} > 0.5$~T or $B^{\|111} > 1.5$~T, whereas intermediate values of $\kappa_{\rm{mag}}$ are observed in the kagom\'{e}-ice phase with $D=3$ for $B^{\|111} < 1$~T and the state for $B^{\|110} > 0.5$~T with fully polarized  $\alpha$ chains along $[110]$ and $\beta$ chains along $[1\bar{1}0]$ with quasi-free spins perpendicular to the field. 

An analogous analysis is more difficult for \hto , because the $\kappa(B)$ curves do not show plateau-like features around 1~T (see Fig~\ref{fig:1}). Concerning the comparison with the data of \hyto , it is also clear that again considering just the difference of both data sets does not yield reliable results. Nevertheless, this comparison reveals that the low-temperature $\kappa(B)$ curves of \hto\ show a sharp low-field decrease suggesting the presence of a sizeable $\kappa_{\rm{mag}}$ in zero-field, whereas the corresponding $\kappa(B)$ curves of \hyto\ only weakly change with field for $B<1$~T (see Figs.~\ref{fig:2} and~\ref{fig:3}). Thus, it appears again reasonable to assume an essentially field-independent $\kappa_{\rm{ph}}^{B\rightarrow 0}$ for \hto\ and in order to get at least a rough estimate of $\kappa_{\rm{mag}}(B)\approx \kappa(B) - \kappa_{\rm{ph}}^{B\rightarrow 0}$ we assume $\kappa_{\rm{ph}}^{B\rightarrow 0} \approx \kappa(B=1\,{\rm T})$. These differences at various fixed temperatures then yield an estimate of the temperature-dependent zero-field $\kappa_{\rm{mag}}(T)$ for \hto , which is compared to the corresponding  $\kappa_{\rm{mag}}(T)$ of \dto\ in Fig.~\ref{fig:4}(a). In addition, the estimated $\kappa_{\rm{mag}}(T)$ of \hto\ from Ref.~\cite{Toews2013} is also displayed, which, as already mentioned above, is much smaller than our results. Note that all three estimates of $\kappa_{\rm{mag}}(T)$ refer to $B=0$, but are measured with different directions of the heat current $j$. With respect to a possible monopole heat transport, the direction of the heat flow should be of minor importance, because an isotropic  monopole mobility can be expected in $B=0$. Moreover, one may also expect that a possible monopole contribution  $\kappa_{\rm{mag}}(T)$ should be of comparable order of magnitude for the two spin-ice materials \hto\ and \dto\ because of the very similar energy scales characterizing their spin-ice behavior. In view of the above-described experimental uncertainties, the comparison of our $\kappa_{\rm{mag}}(T)$ data of both materials essentially confirms these expectations, which is a basic result of this work. Moreover, our data clearly indicate that $\kappa_{\rm{mag}}(T)$ of \hto\ is significantly smaller than $\kappa_{\rm{mag}}(T)$ of \dto. This difference can be naturally explained by an enhanced spin-flip/phonon scattering in \hto, which would simultaneously explain the reduced $\kappa_{\rm{mag}}$ and $\kappa_{\rm{ph}}$ in zero field and the observed increase of $\kappa_{\rm{ph}}(B)$ with increasing $B$. 

Let us finally discuss why the estimate of $\kappa_{\rm{mag}}(T)$ of \hto\ from Ref.~\cite{Toews2013} is so much smaller than ours. In Ref.~\cite{Toews2013}, temperature-dependent measurements of $\kappa(T)$ at constant fields $B=0$, 6, 8, and 10~T were performed and because the $\kappa(T)$ data in the field range between 6 and 10~T are identical, these high-field data were assumed to represent a field-independent background $\kappa_{\rm{ph}}$. Our measurements of  $\kappa(B)$ of the Ho-based materials confirm such a field-independent $\kappa_{\rm{ph}}(B>5\;{\rm T}$), but there is a pronounced field dependence in the intermediate field range around 2~T. This is not only the case for the configuration $j\|B\| [100]$ discussed so far, but also for the configuration $j\|B\| [111]$ studied in Ref.~\cite{Toews2013}, as is shown exemplary in  Fig.~\ref{fig:4}(c--f). For $B\| [111]$, the  $\kappa(B)$ measurements of the Ho(Dy)-based spin ice show additional features up to about 1.5(1)~T,   which are related to the occurrence of the kagom\'{e}-ice phase for this field direction and are absent in the respective $\kappa(B)$ data of the reference materials (for \dto\ see also Ref.~\cite{Kolland2013}). Our data of Fig.~\ref{fig:4}(e,f) clearly show that using the high-field  data $\kappa(T,B>6\;{\rm T})$ as an estimate of $\kappa_{\rm{ph}}^{B\rightarrow 0}$ overestimates this background considerably and causes a drastic underestimate of the corresponding zero-field $\kappa_{\rm{mag}}(T,B=0)$. In Ref.~\cite{Toews2013}, a finite $\kappa_{\rm{mag}}(T,B=0)>0$ is only found for $T< 0.65$~K, because for higher temperature the high-field data of $\kappa(T)$ exceed those in zero field. As can be seen from Figs.~\ref{fig:1} and~\ref{fig:4}, our high-field data of $\kappa$ for both configurations of $B$ and $j$ are larger than the corresponding zero-field data down to our lowest temperature of 0.4~K. Nevertheless, our data agree to those of Ref.~\cite{Toews2013} insofar  that the difference $\kappa(T,B>6\;{\rm T})-\kappa (T,B=0)$ is systematically decreasing with decreasing temperature and one may also suspect a sign change at somewhat lower temperature.

\section{Summary}
\label{sec:Sum}

In conclusion, we observe clear experimental evidence for a sizeable magnetic contribution $\kappa_{\rm{mag}}$ to the low-temperature, zero-field heat transport of both spin-ice materials \hto\ and \dto . We attribute this $\kappa_{\rm{mag}}$ to the magnetic monopole excitations, which are highly mobile in zero field and this mobility is effectively suppressed in external magnetic fields causing a drop of $\kappa_{\rm{mag}}(B)$ in the low-field range. Towards higher magnetic fields, we find significant field dependences of the phononic heat conductivities $\kappa_{\rm{ph}}(B)$ of \hto\ and \dto , which are, however, of opposite signs, as it is also the case in the highly dilute reference materials \hyto\ and \dyto. As discussed earlier~\cite{Kolland2013}, the decreasing  $\kappa_{\rm{ph}}(B)$ in the Dy-based materials probably arise from field-induced lattice distortions, which are seen in magnetostriction data. This effect seems to be less important in the Ho-based materials, which show a significantly smaller magnetostriction while at the same time the scattering of phonons by spin flips appears to be significantly stronger than in the Dy-based materials.  Consequently, both $\kappa_{\rm{mag}}$ and $\kappa_{\rm{ph}}$ in zero field are smaller in \hto\ than they are in \dto\ and the field dependences of $\kappa_{\rm{ph}}$ are of opposite signs.

\section*{Acknowledgement}
This work was supported by the Deutsche Forschungsgemeinschaft through SFB 608 and the project LO 818/2-1.

\end{document}